# Resolution Improvement of the Common Method for Presentating Arbitrary Space Curves Voxel

Al-Oraiqat Anas M. [*1], Bashkov E.A. [2], Zori S.A. [2]

[*1]Taibah University, Department of Computer Sciences and Information
Kingdom of Saudi Arabia, P.O. Box 2898
Email: anas_oraiqat@hotmail.com
[2] SHEE «Donetsk National Technical University», Ukraine, P.O.Box 85300
Email: eabashkov@i.ua, zori@pmi.dgtu.donetsk.ua

**Abstract** - The task of voxel resolution for a space curve in video memory of 3D display is set. Furthermore, an approach solution of voxel resolution of arbitrary space curve, given in parametric form, is studied. Numerous numbers of intensive experiments are conducted and interesting results with significant recommendations are presented.

## I. INTRODUCTION

Three-dimensional (3D) technologies presentation of visual information has been a new area for researchers in the last decade [1-8]. The particular feature of such devices is the presence of a spatial "voxel" memory device. This means that the 3D extension of the "flat" video memory is considered similar to the two-dimensional (2D) display standard devices. The 3D image generation mechanism is accomplished by creating a "voxalized" model of real objects in system memory. The created model consists of 3D graphical primitives such as the intervals of 3D lines, the 3D planes, the arcs, the circumferences, the spherical triangular, the ellipsoids etc.

Studying the development of the approaches for generating voxel presentation of arbitrary space curves and planes is a hot area for research. Actually, in special devices enabling the reproduction effectively is done by reproducing various flat and special splines.

## II. PROPOSED SYSTEM

### A. Task Assignment

The task of generating the voxel presentation of space curve for spatial 3D DSI is presented in [9]. The spatial 3D DSI is presented in the form of parallelepiped projected by a 3D display, the area of 3D Euclidean space.

Considering $\Omega$ as a 3D cube where it is defined as follows:

$$\Omega \in R^3, 0 \leq x \leq X, 0 \leq y \leq Y, 0 \leq z \leq Z.$$

Assuming that the allowances of scaling $X = Y = Z = H$.

$\Omega$ is assumed to be filled with voxels atomic elements which are reflected by a 3D display. Voxel is defined as a cube positioned by axes $\Omega$ with a single rib. Multitude of voxels can be presented as a 3D massive of voxels $V_{i,j,l}$, which is used to fill $\Omega$. The voxel indexes $i,j,l$ are in the massive adopting values $0,1, ..., [H]$. On the other hand, voxel coordinates in $\Omega$ can be defined. Consequently, voxel $V_{i,j,l}$ is sub multitude $\Omega$ which is the corresponding choice $H$. $H$ may be defined as $i \leq x \leq i+(1-\varepsilon)$, $j \leq y \leq j+(1-\varepsilon)$, $cl \leq z \leq l+(1-\varepsilon)$, where $\varepsilon$ is infinitesimal.

For some voxel $V^{(k)}$ with coordinates $i_k, j_k, l_k$ are considered to be neighbours with those voxels $V^{(g)}$, to meet the condition requirement as shown in equation (1).

$$Max\{ |ig-ik|, |jg-jk|, |lg-lk| \}=1 \qquad (1)$$

The function appears in equation (2) is accepted while conducting further speculations in the way of metrics for the multitude of voxels in correspondence with [10].

$$mg,k = |ig,-ik,| + |jg-jk,| + |lg-lk| \qquad (2)$$

The coordinates $V_C$ of the voxel centre $V_{i,j,l}$ in $\Omega$ is defined as shown in equation (3)

$$V_{Cx} = i+0.5, \ V_{Cy} = j+0.5, \ V_{Cz} = l+0.5 \qquad (3)$$

$\Omega$ in some known way is assumed as appears in [11], there is a given curve (univalent, infinite, non-isolated or isolated, one leg) $\Phi(x,y,z)$. This is summarized in figure 1. The curve here has defined initial $S = [x_S, y_S, z_S]$ and final $E = [x_E, y_E, z_E]$ points. For non-isolated curve $E \neq S$ is needed while for isolated curve $E = S$ is accepted. Based on the earlier facts movement direction from $S$ to $E$ is appointed.

It is required to find such voxel consequence $V_{SE}^{(k)}$, $k=1,2,...,N$, that $S \in V_{SE}^{(1)}$, $E \in V_{SE}^{(N)}$, and for any intermediate voxel $V_{SE}^{(k)}$, $k=2,3, ..., N-1$ the condition is met. There is such a point on the curve $\Phi(x,y,z)$ with the following conditions: (a) the length is perpendicular, (b) the starting point is from the centre $V^{(k)}$ onto $\Phi(x,y,z)$, (c) it should not exceed half diagonal of the voxel (for singular voxel 0.866).

The established consequence "3D digital approximation" or "3D screen decomposition" of a space curve and the procedure of acquiring consequence $V^{(k)}$ will be named as 3D generation of a space curve.

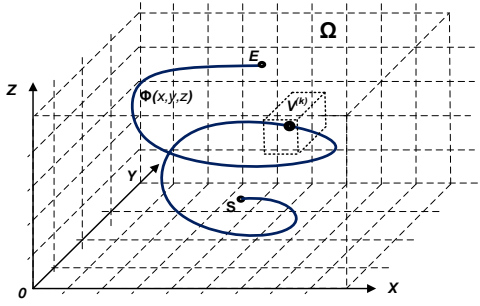

Figure 1 – voxel presentation of arbitrary space curve.

The distance $m_{k,k+1}$ between any consequence pair of voxels in $V_{SE}^{(k)}$ has reached its maximum. To conclude, the neighbour voxels are reduced from its average to be two voxels only. This is summarized in equation (4).

$$m_{k+1,k} = \max_q (m_{q,k}) \quad (4)$$

For all $q$ for which $m_{q,k} \leq 3$.

*B. Method of 3D generation of arbitrary space curve*

The challenge of voxel presentation of arbitrary surfaces and curves in 3D space is closely connected in terms of history. The presentation of flat curves in 2D for generates images on standard 2D reflecting devices. Practically, all known methods for both 2D devices and 3D devices are based on the search of consequence of pixels. 2D and voxels (3D) are normally located at the minimum distance from the assigned curve.

For 2D devices The Bresenhem method is considered to be the best as shown in [12]. Generating images of interval lines based on a comparison of distances to the interval. This is located in the screen plane $OXY$ which is assigned tacitly by equation $Ax + By + C = 0$. This is replaced by evaluation of function value $\varepsilon(x,y) = Ax + By + c$. At this in plane $OXY$ where the interval is located in accordance with increment ability of pixel centres advance on coordinates $x$, $y$ and linearity $\varepsilon(x,y)$. All calculations can be done integrally and exclude multiplication operators. The work in [13-14] extend the Bresenhem method for generation of intervals in 2D space and efforts to use Bresenham approach for curves in 3D space. For line intervals in 3D space modification of Bresenham's algorithm works quite well. For flat curves the substitution of distance estimation for computation $\varepsilon(x,y)$ is not always fair. Almost always estimation $\varepsilon(x,y)$ can't be performed integrally.

The work in [15-20], the algorithms for voxel decomposition for various special surfaces and curves has been studied. The Prescribed in the evident form, as well as algorithms of special arc generation have been suggested [19- 20]. The essence of the algorithm in [20] lies in the following:

$V_{SE}^{(k)}$ is defined to be $k$- voxel of the required consequence. In the point of the curve, nearest to $V_{SE}^{(k)}$, there is a tangent line to curve $\vec{G}$ and the matrix of advance $M^{(k)}$ is summarized as follows:

$$\begin{bmatrix} sign(G_x) & 0 & 0 & sign(G_x) & 0 & sign(G_x)sign(G_x) \\ 0 & sign(G_y) & 0 & sign(G_y) & sign(G_y) & 0 & sign(G_y) \\ 0 & 0 & sign(G_z) & 0 & sign(G_z) & sign(G_z)sign(G_z) \end{bmatrix}$$

where $G_x, G_y, G_z$ coordinate vector of tangent $\vec{G}$ to a curve in point $V_{SE}^{(k)}$. Matrix $M^{(k)}$ defines 7 trial voxels – challengers $V_{SE_q}^{(k+1)} = V_{SE}^{(k)} + M_q^{(k)}$, $q = 0,1,\ldots 6$, among which in terms of minimum distance to a curve the next $(k+1)$ consequence voxel is chosen.

In [19-20], considering the implicit assignment of a circle where targeted arc belongs, and the attached arbitrarily-spaced plane for computing the tangent to the arc have used particular specific algorithms for the required curve requirement. It is impossible to apply such algorithms for searching a tangent to an arbitrary space curve.

If a space curve is assigned in parametric form

$$x(\varphi), y(\varphi), z(\varphi), \quad \varphi_{min} \leq \varphi \leq \varphi_{max},$$

where $\varphi$ is a parameter assigning the position of a space curve point in 3D space, the vector of tangent $\vec{G}$ of a curve is defined as as shown in equation (5)

$$G_x(\varphi) = \frac{dx(\varphi)}{d\varphi}, \quad G_y(\varphi) = \frac{dy(\varphi)}{d\varphi}, \quad G_z(\varphi) = \frac{dz(\varphi)}{d\varphi}. \quad (5)$$

Hence, assigning a curve in a parametric form enables estimating the tangent coordinates in the assigned point easily. If it is impossible to find derivatives analytically, tangent computation is limited to formula evaluation. Otherwise it is allowed to apply some difference approximation.

Taking into account the above mentioned considerations, the following algorithm of voxalization of an arbitrary space curve is proposed assigned in the parametric form. The voxel decomposition algorithm of a spherical triangular on pseudocode is summarized as follows:

**start**
    assigning arbitrary curve
    assigning initial $S$ and final $E$ points;
    $V_{SE}^{(k)} = S$
    **while** $V_{SE}^{(k)}$ doesn't include **E**
        computation for component of tangent $\vec{G}$ for $V_{SE}^{(k)}$
        forming $M^{(k)}$
        **for** $q:=0, 1,\ldots 6$ **do**
            $V_{SE_q}^{(k+1)} = V_{SE}^{(k)} + M_q^{(k)}$
            calculation $Dist_q^{(k+1)}$
            $q^* = arg\,(\min_q Dist_q^{(k+1)})$
            $V_{SE}^{(k)} = V_{SE_{q^*}}^{(k+1)}$
        **endfor**
    **endwhile**
**finish**

For the efficiency of the given algorithm it is required among $V_{SE_q}^{(k+1)}$ to find at least one voxel, which the distance $Dist_q^{(k+1)}$) from its centre to the curve doesn't exceed 0.866. In most cases, it imposes previously unknown limitation for maximum allowable value of first and second curvature.

An important issue concerning the efficiency of the algorithm lies in the following: in what way and how much accurately it is required to evaluate the distance from $V_{SE_q}^{(k+1)}$ to the nearest point of a curve in case the following is chosen as shown in equation (6)

$$V_{SE}^{(k+1)} = V_{SE_{q*}}^{(k+1)}, . \qquad (6)$$

where $q^*$ is such that

$$q^* = arg(\min_q Dist_q^{(k+1)}), \quad q = 0,1,2,\ldots,6$$

## III. EXPERIMENTAL STUDY OF THE VOXEL DECOMPOSITION ALGORITHM OF AN ARBITRARY SPACE CURVE.

Based on the experiments, the estimated maximum errors of the received voxel consequence is:

$$\varepsilon_{max} = \max_k \left(Dist_{q*}^{(k)}\right), \quad k = 1,2,\ldots N \qquad (7)$$

and average

$$\varepsilon_{av} = \frac{1}{N}\sum_{k=1}^{N} Dist_{q*}^{(k)}, \quad k = 1,2,\ldots N \qquad (8)$$

At $\varepsilon_{max} < 0.866$ generated approximation doesn't break the conditions of attributing some point of a curve to each voxel of the consequence, and $\varepsilon_{av}$ characterizes the quality of approximation.

Experiments were conducted for cylindrical space curve, formed by the movement of the point around the cylinder of radius $R$, and parallel to axis $0Z$. Central axis of which is dislocated in the plane $X0Y$ to $x_0, y_0$, are related to origin of coordinates. In coordinate-parametrical form the curve is assigned as shown in equation (9):

$$\begin{cases} x(\varphi) = x_0 + R\cos\varphi, \\ y(\varphi) = y_0 + R\sin\varphi, \\ z(\varphi) = z_0 + A(1 - \cos(\omega\varphi)), \\ \varphi_{min} \leq \varphi \leq \varphi_{max} \end{cases} \qquad (9)$$

where: $R$- radius of cylindrical surface; $x_0, y_0$ - dislocation of the main axis of the cylindrical surface in plane X0Y relative to origin of coordinates; $A$- Amplitude of point oscillation along the cylinder guide; $\omega$– Point oscillation frequency along the cylinder guide.

Tangent vector in this case equals as shown in equation (10):

$$\begin{cases} G_x(\varphi) = -R\sin(\varphi) \\ G_y(\varphi) = R\cos(\varphi) \\ G_z(\varphi) = \omega A\sin(\omega\varphi) \end{cases} \qquad (10)$$

The curvature of a curve is regulated by the change of frequency $\omega$ of point oscillation along axis $0Z$. During the experiments there varied both the procedure of tangent computation and the algorithm of distance computation.

Estimating the tangent $\vec{G}$ were used either analytically acquired expressions of a component, or approximation of derivatives by the first right difference.

Calculation of the distance $Dist_q^{(k+1)}$ was performed in three ways.

*Firstly.* For arbitrary voxel $V^*$ the parameter is calculated

$$\varphi^* = arctg\left(\frac{V_y^*}{V_x^*}\right) \qquad (11)$$

Then curve point $P^* = [x(\varphi^*), y(\varphi^*), z(\varphi^*)]$ is defined and further the distance $Dist$ between centres $V^*$ and $P^*$.

*Secondly.* The angle $\varphi^*$ is defined similarly to step one which used equation (11), and then for $\varphi_{test}$ changing from $\varphi^* - \varphi_d$ до $\varphi^* + \varphi_d$ with some low pitch $\Delta\varphi$ there are defined distances $Dist(\varphi_{test})$ от $V^*$ and $P(\varphi_{test})$ and the minimum is chosen.

*Thirdly.* Based on second step, three consequential parameter meanings $\varphi_{test1}, \varphi_{test2}, \varphi_{test3}$, are defined, for which the conditions $Dist(\varphi_{test1}) > Dist(\varphi_{test2})$ and $Dist(\varphi_{test3}) > Dist(\varphi_{test2})$, are followed, and using the parabolic interpolation formulas in [21], the minimum distance is specified.

It should be noted that the topic of the research was voxel approximation error; temporary characteristics of the realized algorithm were not taken into account. There was modelled a process of generation of assigned curve for the display device with resolution 128x128x128 = 2 Mvoxels) and 256x256x256 = 8 Mvoxels).

Curve characteristics are given in table I.

Table I. Characteristics of the curve under study.

| Resolution | 2 Mvox | 8 Mvox |
|---|---|---|
| Radius of a cylindrical surface $R$ | 40 | 80 |
| Amplitude of point oscillation along the directing $A$ | 40 | 80 |
| Offset $x_0 = y_0 = z_0$ | 50 | 100 |
| Change limits of curve parameters $-10º \leq \varphi \leq 10º$ | | |

At the same time, each step of estimating the tangent and calculating the distance point oscillation frequency along the guiding $\omega$ was taken as 2 Hz and 4Hz. Figure 2 shows an example of typical image of generated voxel consequence.

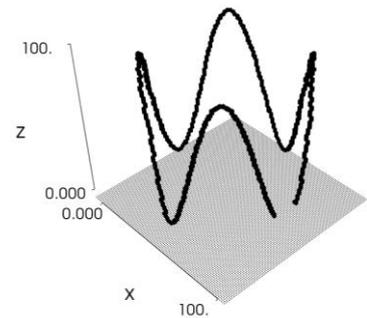

Figure 2-An example of voxel presentation of a curve for resolution 2 Mvox and $\omega = 4Hz$.

Table II summarizes the comparative results of experiments for different procedure of calculating the tangent.

Table II. Computation variants comparison $\vec{G}$ ($\omega = 2\ Hz$).
A: Analytically, Q: Quantitatively

| Estimating the tangent | A | Q | A | Q |
|---|---|---|---|---|
| Resolution | 2 Mvox | | 8 Mvox | |
| Number of voxels | 660 | 660 | 1348 | 1348 |
| $\varepsilon_{max}$ | 0.837 | 0.837 | **1.845** | **1.845** |
| $\varepsilon_{av}$ | 0.501 | 0.501 | 0.803 | 0.843 |

Calculating the distance from voxel to tangent is summarized in Tables III and Table IV.

Table III. Comparison of computation variants $Dist$ for $\omega = 2\ Hz$

| Estimating the tangent | Variant 1 | | Variant 2 | | Variant 3 | |
|---|---|---|---|---|---|---|
| Resolution | 2 M vox | 8 M vox | 2 M vox | 8 M vox | 2 M vox | 8 M vox |
| Number of voxels | 355 | 712 | 347 | 708 | 347 | 700 |
| $\varepsilon_{max}$ | 0.633 | **0.868** | 0.601 | 0.664 | 0.586 | 0.616 |
| $\varepsilon_{av}$ | 0.394 | 0.473 | 0.358 | 0.391 | 0.350 | 0.391 |

Table IV. Comparison of computation variants $Dist$ for $\omega = 4\ Hz$

| Estimating the tangent | Variant 1 | | Variant 2 | | Variant 3 | |
|---|---|---|---|---|---|---|
| Resolution | 2 M vox | 8 M vox | 2 M vox | 8 M vox | 2 M vox | 8 M vox |
| Number of voxels | 660 | 1348 | 644 | 1325 | 651 | 1321 |
| $\varepsilon_{max}$ | 0.837 | **1.845** | 0.669 | 0.812 | 0.654 | 0.795 |
| $\varepsilon_{av}$ | 0.501 | 0.843 | 0.411 | 0.494 | 0.395 | 0.462 |

The analysis of the received experimental data allows gives the following conclusions. The appliance of numerical method for tangent component doesn't cause the loss of algorithm efficiency. However, it leads to some decrease of approximation quality.

Appliance of more accurate algorithms for calculating the distance between a curve and an arbitrary voxel in all cases decreases average error. This indicates to a higher quality of approximation. Besides, experiment results show increasing local curvature (second curvature) of the curve and the increase of device resolution 3D algorithm may lose efficiency (maximum error exceeds 0.866).

## IV. CONCLUSION

Modifications of algorithm for voxel decomposition of arbitrary space curves are aimed by increasing the accuracy of estimating the distance between voxel and the assigned curve. This allows increasing the approximation "quality" perceived as the closeness of approximating consequence of voxels and the assigned curve.

Further studies are conducted in two directions. Firstly, the adjacency of two and only two voxels of the generated consequence is to be guaranteed Secondly, preliminary estimation and automatic account of maximum curvature and second curvature during generation should be considered. This should take into consideration the aim of receiving proper quality of approximation for "ramp" curves.

## V. ACKNOWLEDGMENT

The authors would like to thank Taibah University and Donetsk National Technical University for supporting this research.

## AUTHOR'S PROFILE

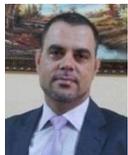
**Anas Mahmoud Al-Oraiqat** received a B.S. in Computer Engineering and M.S. in Computer Systems & Networks from Kirovograd Technology University in 2003 and 2004, respectively, and Ph.D. in Computer Systems & Components from Donetsk National Technical University (Ukraine) in 2011. He has been an Assistant Professor at the Computer & Information Sciences Dept., Taibah University (Kingdom of Saudi Arabia) since Aug. 2012. Prior to his academic career, he was a Network Manager at the Arab Bank (Jordan), 2011-2012. Also, he was a Computer Networks Trainer at Khwarizmi College (Jordan), 2005-2007.

His research interest is in the areas of computer graphics, image/video processing, 3D devices, modelling and simulation of dynamic systems, and simulation of parallel systems.

E-mail: anas_oraiqat@hotmail.com

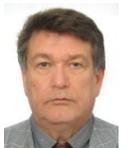
**Evgeniy A. Bashkov** received a DrSci in Computer Systems & Components from Pukhov Institute for Modeling in Energy Engineering National Academy of Science of Ukraine in 1995 and Professor in 1996. He is Professor of Applied Mathematics & Computer Science Department of the Faculty of Computer Science and Technologies of Donetsk National Technical University ("DonNTU").

His research interest is in the areas of the real time computer graphics, virtual reality, the computer image generators, the simulation system, the high performance computing systems of common and special purpose.

E-mail: eab23may@gmail.com

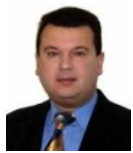
**Zori Sergii A.** received a Ph.D. in Computer Systems & Components from NTU "Donetsk National Technical University" (Ukraine) in 1996 and Associated Professor in 1997. He is Associate Professor of Applied Mathematics & Computer Science Department of the Faculty of Computer Science and Technologies of Donetsk National Technical University ("DonNTU").

His research interest is in the areas of computer graphics, image/video processing & vision, 3D devices, Virtual reality systems, specialized parallel computer systems.

E-mail: sa.zori1968@gmail.com